\documentclass[twocolumn,showpacs,preprintnumbers,draftclsnofoot]{revtex4}
\usepackage{amsmath}
\usepackage{amssymb}
\usepackage{graphicx}
\usepackage{dcolumn}
\usepackage{bm}
\usepackage{amssymb}
\usepackage{graphicx}
\usepackage{dcolumn}
\usepackage{bm}
\begin{document}

\title{ Coherent Goos-H$\ddot{a}$nchen shifts of meta-grating with radiation asymmetry }
\author{Ma Luo\footnote{Corresponding author:swym231@163.com} and Feng Wu }
\affiliation{School of Optoelectronic Engineering, Guangdong Polytechnic Normal University, Guangzhou 510665, China}

\begin{abstract}

The coherent Goos-H$\ddot{a}$nchen shifts of meta-grating are proposed, which is the Goos-H$\ddot{a}$nchen shifts of the two outgoing beams under the simultaneous incidence of two coherent optical beams from opposite sides of the grating with the same lateral wave number. As both of the frequency and lateral wave number are resonant with a topological state of the meta-grating, such as unidirectionally guided resonance or circular polarized states, the energy flux and Goos-H$\ddot{a}$nchen shifts of the two outgoing beams are coherently controlled by the relative phase difference between the two incident beams. By applying stationary-phase method, it is found that the enhancement of coherent Goos-H$\ddot{a}$nchen shifts by the unidirectionally guided resonance and circular polarized states is accompanied by constant and peak transmittance, respectively. Analysis with temporal coupled mode theory shows that the different features are due to difference mechanism of interference between direction scattering and resonant radiation. The coherent Goos-H$\ddot{a}$nchen shifts with incident Gaussian beams are sensitive to the relative phase between the two beams, which can be applied in refractive index sensor.

\end{abstract}

\maketitle

\section{Introduction}

At an optical interface, the reflected or transmitted optical beam could experiences a lateral shift relative to the oblique incident optical beam, which is called Goos-H$\ddot{a}$nchen (GH) shift \cite{firstGH47}. The GH shift can be enhanced by exciting varying types of localized resonant state of the meta-surface at the interface, such as interface with Brewster effects \cite{Brewster1,Brewster2,Brewster3}, surface plasmon polariton \cite{sppGH1,sppGH2,sppGH3,sppGH4}, Fabry-Perot resonances \cite{fpcavityGH1,fpcavityGH2,fpcavityGH3,fpcavityGH4}, Bloch surface waves \cite{blochGH1,blochGH2}, Tamm plasmon polaritons \cite{tamnGH1,tamnGH2,tamnGH3}, and quasi-bound states in the continuum (BIC) \cite{bicGH1,bicGH2,bicGH3}. The GH shift can be theoretically explained by the stationary-phase method \cite{stationaryPt48}, which predicts the GH shift by calculating the slope of reflection (transmission) phase versus incident angle. The GH shift of reflected (transmitted) optical beams is highly dependent on the reflectance (transmittance) and reflection (transmission) phase. In order to enhance GH shift, large slope of reflection (transmission) phase needs to coincide with the peak of the reflectance (transmittance) \cite{bicGH1}. By tuning the structural parameters or refractive index, the GH shift can be modified. Thus, a scheme of refractive-index sensing based on GH shift has been proposed \cite{bicGH3}.

Asymmetric bilayer all-dielectric meta-grating hosts varying types of topological resonant states, such as BIC, unidirectionally guided resonant (UGR), and circular polarized states (CPS) \cite{raDongJianWen}. BICs in photonic systems are singular states above the light cone, but without radiative loss \cite{bicItSelf1,bicItSelf2,bicItSelf3,bicItSelf4}. BICs usually locate in the center of a vortex in a parameter space, where far-field radiative coefficients are strictly equal to zero \cite{bicVortex}. The dimensions of the parameter space include structural parameters and optical field parameters such as lateral wave number $k_{x}$ \cite{raDongJianWen,bicVortex}. When a parameter is tuned away from the vortex center, the BIC is transferred into quasi-BIC with finite but ultra-high Q factor \cite{quasiBIC01,quasiBIC02,quasiBIC03,quasiBIC04,quasiBIC05,quasiBIC06,quasiBIC07,quasiBIC08,quasiBIC09,quasiBIC10,quasiBIC11,quasiBIC12,quasiBIC13,quasiBIC14,quasiBIC15,quasiBIC16,quasiBIC17,quasiBIC18,quasiBIC19,quasiBIC20,quasiBIC21}. Incidence of an optical field to the grating with the same parameters (both structural and optical parameters) as the quasi-BIC can induce large local field enhancement at the meta-grating. If the symmetry of the system is broken by tuning the parameters, the vortex could be split into two vortices with half topological charge, such as C point with CPS \cite{raDongJianWen,cPointFromV1,cPointFromV2,cPointFromV3}. For a CPS, the radiative loss to up and down sides of the grating has equal magnitude, but with $\delta=\pm\pi/2$ phase difference. Incidence of two coherent optical fields to the grating with the same parameters as the CPS from opposite sides can also induce local field enhancement. When the phase difference between the two optical fields (designated as $\varphi$) equates to $\delta$ or $\delta+\pi$, the local field enhancement reaches maximum and minimum, respectively. An UGR is in the center of vortex of the far-field radiation to either up or down side of the grating, so that the UGR has radiative loss to only down or up side, respectively \cite{cPointFromV1,cPointFromV3}. Incidence of an optical field to the grating with the same parameters as the UGR from the side with or without radiative loss can induce large and small local field enhancement. Large local field enhancement implies strong excitation of the corresponding resonant state, which could enhance various type of optical effect, such as GH shift \cite{bicGH1}, coherent perfect absorption \cite{raDongJianWen}, and second harmonic generation \cite{bicSHG1,bicSHG2}.

\begin{figure}[tbp]
\scalebox{0.19}{\includegraphics{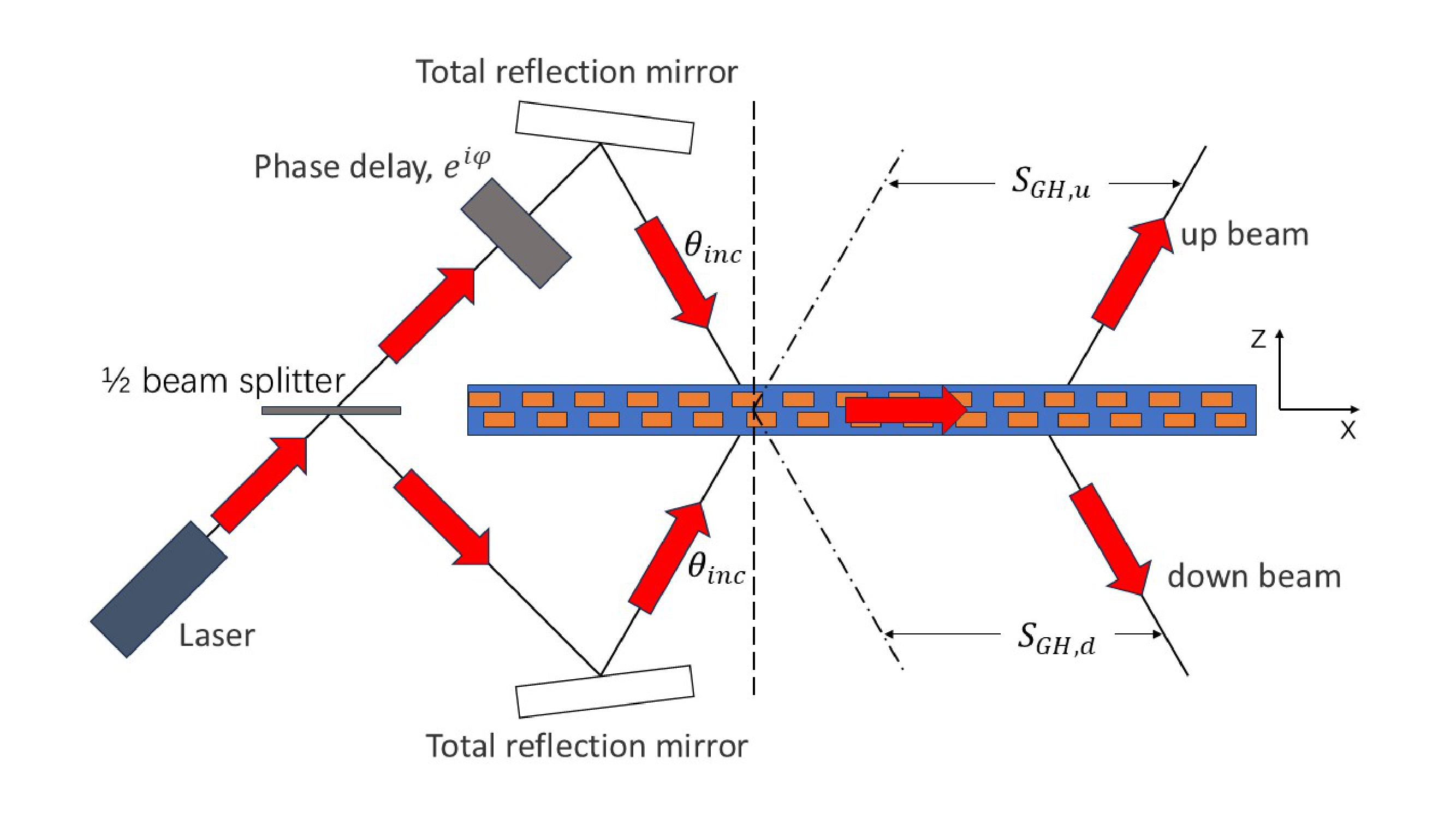}}
\caption{ The scheme of coherent Goos-H$\ddot{a}$nchen shift. An optical beam is split into two beams by 1/2 beam splitter. One of the beams has phase delay $e^{i\varphi}$. Two total reflective mirrors redirect the two beams, which irradiate the meta-grating with the same incident angle from opposite sides of the grating, but the same side of the normal line. The outgoing beams to the top and bottom background are designated as R beam and T beam, whose Goos-H$\ddot{a}$nchen shifts are designated as $S_{GH,r}$ and $S_{GH,t}$, respectively.     }
\label{figure_scheme}
\end{figure}

In this paper, we proposed a set of photonic systems, which uses the phase difference between two incident optical beams (i.e., $\varphi$) to manipulate the excitation of topological resonant states, such as UGR and CPS, which in turn tunes the GH shift of the two out-going optical beams, as shown in Fig. \ref{figure_scheme}. The scheme is designated as coherent GH shift. An optical beam from a continuous-wave laser source with output power being $P_{inc}$ is split into two coherent optical beams with the same magnitude. Two mirrors redirect the two optical beams, which are incident onto a meta-grating from the up and down sides of the grating with the same incident angle to the normal line. Thus, the two optical beams have the same horizontal component of the wave number, i.e., the same $k_{x}$, and opposite $k_{z}$. One of the optical beams has an extra phase $e^{i\varphi}$, due to phase-delay optical component, so that the two optical beams have phase difference $\varphi$. Because the two optical beams are coherent with each other, $\varphi$ is a definite number. After being scattered by the meta-grating, the outgoing beams to the up and down sides of the grating have lateral shifts from the incident point, which are the corresponding GH shifts designated as $S_{GH,u}$ and $S_{GH,d}$, respectively. Because $S_{GH,u(d)}$ are dependent on $\varphi$, the GH shift can be coherently controlled by the phase difference between the two optical beams, so that the effect is designated as coherent GH shift. If the frequency and incident angle of the incident beams are resonant with a quasi-BIC, the magnitude of the GH shifts is highly enhanced. If the quasi-BIC has exotic topological feature, such as UGR and CPS, the GH shifts are strongly sensitive to $\varphi$.

The remainder of this paper is organized as follows. In Sec. II, the branches of UGR and CPS as subsets of quasi-BICs is studied. In Sec. III, the GH shift of the UGR and CPS are studied by stationary-phase method and numerical simulation of incidence of a Gaussian beam with finite beam width. In Sec. IV, conclusions are presented.

\section{Branches of UGR and CPS as subset of quasi-BIC}

\begin{figure*}[tbp]
\scalebox{0.53}{\includegraphics{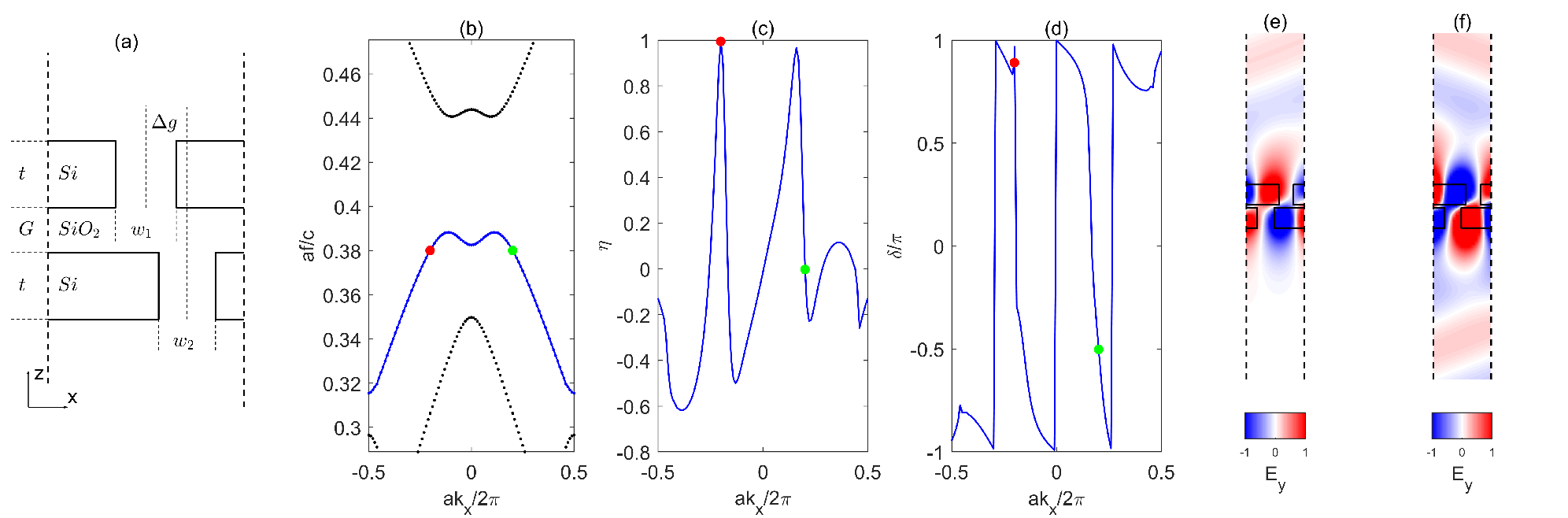}}
\caption{ (a) The structure of the double layer dielectric meta-grating and the corresponding structural parameters. (b) The band structure of the TE mode for a meta-grating with $t=0.355a$, $G=0.045a$, $w_{1}=0.25a$, $w_{2}=0.296a$, and $\Delta p=0.354a$. The directionality $\eta$ and the phase difference $\delta$ of the blue bands are plotted in panels (c) and (d), respectively. In the panels (b-d), the red and green dots marks the UGR and CPS, respectively. The field pattern of the two modes at the blue band with $ak_{x}/2\pi=-0.201$ and $+0.201$ are plotted in the panels (e) and (f), respectively. }
\label{figure_bandUGR}
\end{figure*}

By engineering the structure parameters of a double-layer meta-grating, the optical resonant mode with varying type of topological features can be obtained, such as BIC, UGR, and CPS. The structure of the double-layer meta-grating is plotted in Fig. \ref{figure_bandUGR}(a). Two Si slabs with thickness being $t$ are stacked along the z direction with an interval $G$. The top and bottom slabs are periodically carved with slits along the x direction with period being $a$, and width being $w_{1}$ and $w_{2}$, respectively. The center of the slits in the top and bottom layers have a lateral offset $\Delta p$. The background medium is SiO$_{2}$. The refractive index of Si and SiO$_{2}$ is assumed to be $n_{g}=3.4767$ and $n_{b}=1.444$, respectively. The meta-grating hosts optical resonant modes with Bloch periodic boundary condition, so that each resonant mode in the band structure is marked by the Bloch wave number $k_{x}$ and resonant frequency $f$. Because $k_{x}$ and $\Delta p$ are periodic with period being $2\pi/a$ and $a$, respectively, they form a synthetic parameter space. In this paper, we studied varying structures with the same value of $a$, $t$, and $G$, but varying values of $w_{1}$, $w_{2}$, and $\Delta p$. Thus, the structural parameter space $(w_{1},w_{2})$ and the synthetic parameter space $(k_{x},\Delta p)$ form a four dimensional parameter space. The branches of UGR and CPS are found in the four-dimensional parameter space.

For the case with $w_{1}=w_{2}=0$, the grating degrades into a dielectric slab with waveguide modes. By applying the Bloch periodic boundary condition, the waveguide modes are folded into the first Brillouin zone and form multiple band crossings. As $w_{1(2)}$ become nonzero, the waveguide modes are perturbed, and the two modes at each band crossing couple. The coupling induces avoided band crossing with band gap. Varying types of topological optical resonant modes could be induced by the coupling between two optical modes. The far field pattern of a TE mode with resonant frequency $f$ and lateral wave number $k_{x}$ can be written as
\begin{equation}
E_{y}=\{\begin{array}{cc}
c_{up}e^{ik_{x}x+ik_{z}} & z\rightarrow+\infty  \\
c_{down}e^{ik_{x}x-ik_{z}} & z\rightarrow-\infty \\
\end{array}
\end{equation}
with $k_{z}=\sqrt{n_{b}^{2}k_{0}-k_{x}^{2}}$, $k_{0}=2\pi f/c$ and $c$ being speed of light. The BIC is featured by $c_{up}=c_{down}=0$; the UGR is featured by $c_{up}=0$ or $c_{down}=0$; the CPS is featured by $c_{up}/c_{down}=\pm1i$. The far-field radiative coefficients $c_{up(down)}$ can be extracted from the near field pattern of the eigenstate $E_{y}$ by performing Fourier transform on the up and down boundary of the computational domain. The field pattern $E_{y}$ within one unit cell is calculated by solving the Helmholtz wave equation with Finite Element Method (FEM) \cite{theFEM1,theFEM2,theFEM3,theFEM4,theFEM5,theFEM6}. The Bloch periodic boundary condition is applied to the left and right boundaries of the computational domain. Perfectly Matched Layer (PML) is added to the up and down boundaries of the computational domain.

In order to characterize the topological optical resonant modes with more concise notation, we define directionality as  $\eta=(|c_{up}|^{2}-|c_{down}|^{2})/(|c_{up}|^{2}+|c_{down}|^{2})$, and phase difference as $\delta=arg(c_{up}/c_{down})$. Thus, the UGR is featured by $\eta=\pm1$; the CPS is featured by $\eta=0$ and $\delta=\pm\pi/2$. By further defining a radiation asymmetry pseudopolarization, given as $\vec{c}=\frac{1}{N}[(c_{up}+c_{down})\vec{e}_{1}+(c_{up}-c_{down})\vec{e}_{2}]$, with $\vec{e}_{1}$ and $\vec{e}_{2}$ being a pair of orthogonal basis, and $N=\sqrt{|c_{up}+c_{down}|^{2}+|c_{up}-c_{down}|^{2}}$ being the normalization coefficient, the BIC and CPS can be characterized by a vortex in the synthetic parameter space with topological charge being $\pm1$ and $\pm1/2$, respectively. When $w_{1}=w_{2}$, the BIC is protected by the central inversion symmetry, so that the topological charge is $\pm1$. When $w_{1}\ne w_{2}$, the central inversion symmetry is broken, so that the vortex with topological charge $\pm1$ is split into two C points with topological charge $\pm1/2$.

A specific structure without central inversion symmetry that hosts both UGR and CPS in the same band structure is found. The corresponding band structure is plotted in Fig. \ref{figure_bandUGR}(b), with the structural parameters being given in the caption of the figure. The blue band in Fig. \ref{figure_bandUGR}(b) has a large avoided band crossing with the other bands. The directionality $\eta$ and the phase difference $\delta$ of the blue bands are plotted in Fig. \ref{figure_bandUGR}(c) and (d), respectively. As $ak_{x}/2\pi=-0.201$, the directionality is equal to $1$, which features a UGR without radiative loss to the down side of the meta-grating. The corresponding field pattern of $E_{y}$ is plotted in Fig. \ref{figure_bandUGR}(e), which shows that plane wave carrying radiative loss only travels to the up side of the meta-grating. The angle between the direction of the traveling plane wave and the normal line is given as $\sin^{-1}[|k_{x}|/(n_{b}k_{0})]$. On the other hand, as $ak_{x}/2\pi=+0.201$, the directionality is equal to 0, and the phase difference is equal to $-\pi/2$, which feature a CPS. The corresponding field pattern of $E_{y}$ is plotted in Fig. \ref{figure_bandUGR}(f), which shows that the plane waves carrying radiative loss to the up and down sides have $-\pi/2$ phase difference, and the same magnitude.

\begin{figure}[tbp]
\scalebox{0.58}{\includegraphics{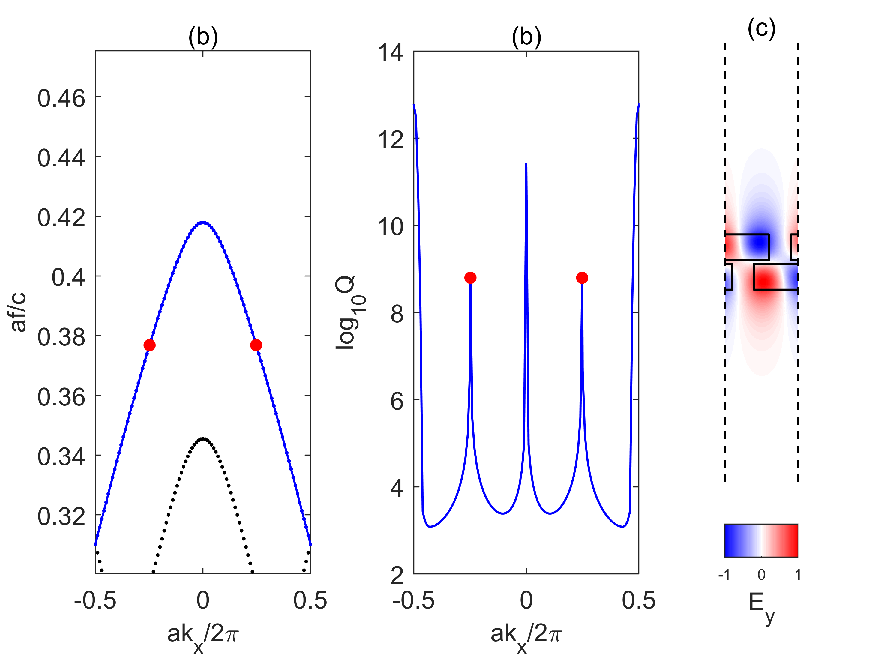}}
\caption{ (a) The band structure of the TE mode for a meta-grating with $t=0.355a$, $G=0.045a$, $w_{1}=w_{2}=0.296a$, and $\Delta p=0.5a$. The Q factor of the blue band is plotted in the panel (b). The two red dots mark the two BICs at $ak_{x}/2\pi=\pm0.248$ that are protected by the central inversion symmetry. (c) The field pattern of the BIC at $ak_{x}/2\pi=0.248$.  }
\label{figure_bandBIC}
\end{figure}

For another specific structure with central inversion symmetry, the band structure is plotted in Fig. \ref{figure_bandBIC}(a). The Q factor of the states of the blue band in Fig. \ref{figure_bandBIC}(a) is plotted in Fig. \ref{figure_bandBIC}(b). The structural parameters are given in the caption of Fig. \ref{figure_bandBIC}. The Q factor tends to be infinite at $k_{x}=0$, $ak_{x}/2\pi=\pm0.248$ and $ak_{x}/2\pi=\pm0.5$. As $k_{x}=0$, the BIC is protected by left-right mirror symmetry, which is a typical symmetry-protected BIC. As $ak_{x}/2\pi=\pm0.248$, the BIC is protected by central inversion symmetry. The corresponding field pattern is plotted in Fig. \ref{figure_bandBIC}(c), which is neither an even nor odd function of the x coordinate, so that there is no symmetry-mismatch with the radiative plane wave. However, the BIC is in the middle of a vortex of the radiation asymmetry pseudopolarization, which guarantees zero magnitude of both $c_{up}$ and $c_{down}$. Thus, this resonant mode is a BIC. As $ak_{x}/2\pi=\pm0.5$, the resonant mode is under the light cone of the SiO$_{2}$ background, so that the mode is a typical waveguide mode.

\begin{figure}[tbp]
\scalebox{0.6}{\includegraphics{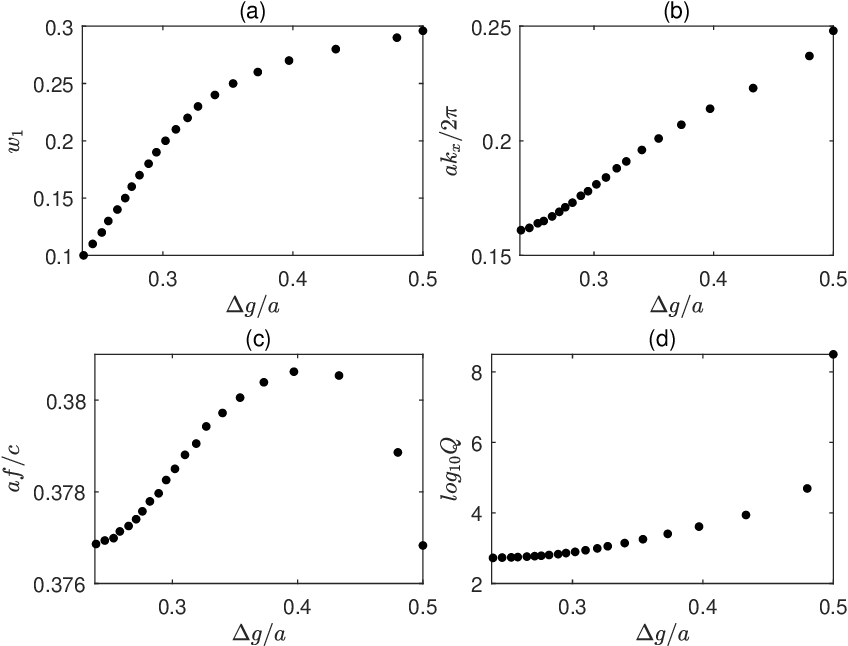}}
\caption{ For the branches of systems with both UGR and CPS in one band, and $w_{2}/a$ being fixed to $0.296$, the structural parameter $w_{1}$ and synthetic parameter $k_{x}$ versus the synthetic parameter $\Delta p$ are plotted in panels (a) and (b), respectively. In the branches of systems, the resonant frequency and Q factor are plotted in panels (c) and (d), respectively.   }
\label{figure_paraUGR}
\end{figure}

As any parameters in the four-dimensional parameter space $(w_{1},w_{2},k_{x},\Delta p)$ change, the BIC in Fig. \ref{figure_bandBIC} becomes a quasi-BIC with finite Q factor. In  other words, $c_{up}$ or $c_{down}$ of the field pattern become nonzero. A subset of the quasi-BIC given by simultaneously tuning the four parameters with a specific combination is the topological systems with both UGR and CPS, which includes the system in Fig. \ref{figure_bandUGR}. Numerical results show that the subset forms a two-dimensional surface in the four-dimensional parameter space. In other words, as either one of the two structural parameters $w_{1,2}$ changes, the two synthetic parameters $\Delta p$ and $k_{x}$ within the subset are functions of the structural parameters, i.e., $\Delta p(w_{1},w_{2})$ and $k_{x}(w_{1},w_{2})$. In order to exhibit the subset, by fixing $w_{2}$ to be $0.296a$, and scaling $w_{1}$ from $0.29a$ to $0.1a$ with an interval of $-0.01a$, we numerically found $\Delta p$ and $k_{x}$, whose corresponding systems have both UGR and CPS. Mathematically, one can switch the independent variables to be $w_{2}$ and $\Delta p$, so that $w_{1}$ and $k_{x}$ within the subset are functions of $(w_{2},\Delta p)$. The $w_{1}$ and $k_{x}$ versus $\Delta p$ with $w_{2}$ being fixed to be $0.296a$ are plotted in Fig. \ref{figure_paraUGR}(a) and (b), respectively. In each system, UGR and CPS are at the same band with opposite $k_{x}$, so that the resonant frequencies (both real and imaginary parts) of the UGR and CPS are the same due to time-reversal symmetry. The resonant frequency and Q factor of the corresponding UGR(CPS) in the series of systems of Fig. \ref{figure_paraUGR}(a,b) are plotted in Fig. \ref{figure_paraUGR}(c) and (d), respectively. The resonant frequency slightly depends on $\Delta p$. As $\Delta p$ approaches $0.5a$, the system regains central inversion symmetry, and the Q factor exponentially approaches infinite.

\section{Coherent GH shift}

\subsection{Plane wave incidence}

\begin{figure*}[tbp]
\scalebox{0.53}{\includegraphics{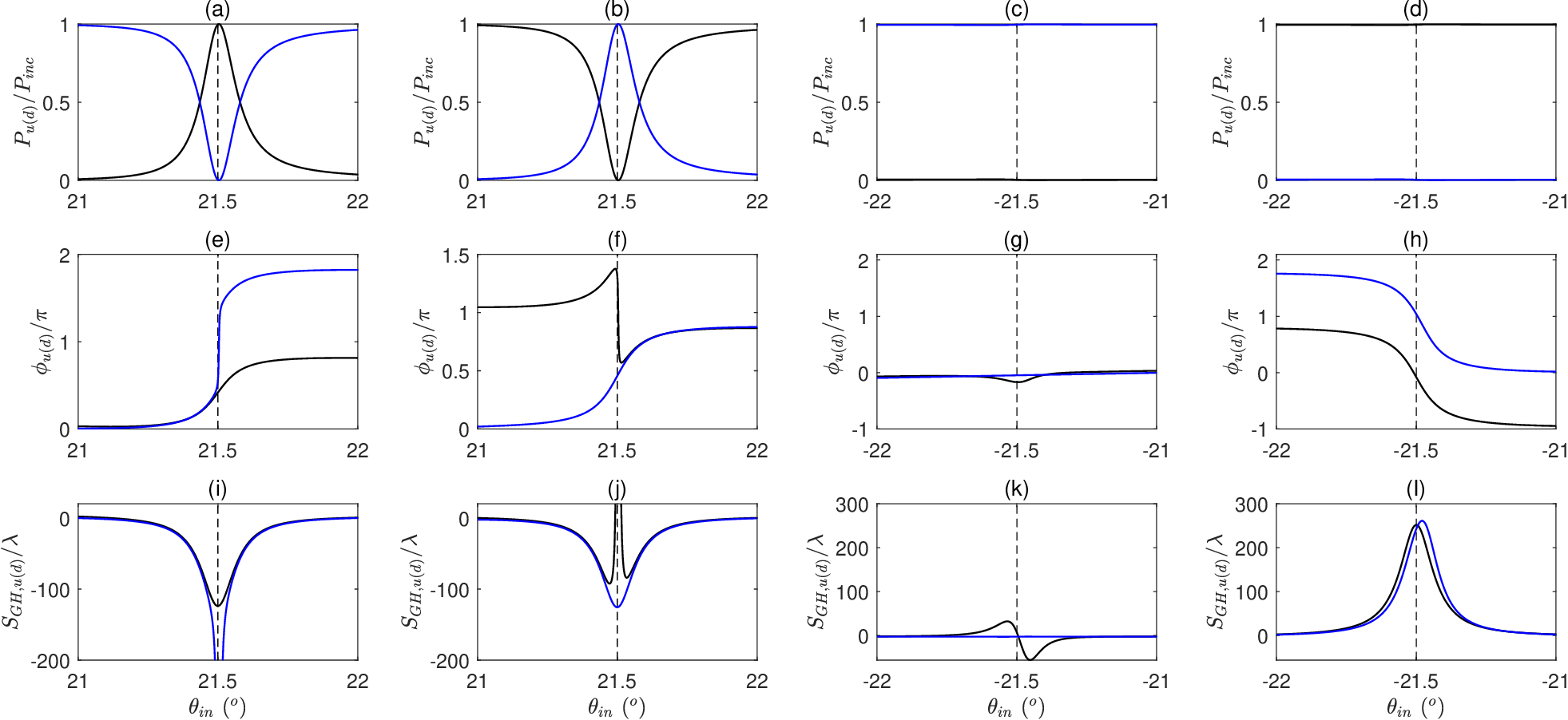}}
\caption{ The angular spectrum of the ratio between the outgoing power to the up and down sides and the total incident power, i.e., $P_{u}/P_{inc}$ and $P_{d}/P_{inc}$ versus incident angle $\theta_{in}$, is plotted as black and blue lines in the first row. The phase of the outgoing plane wave to the up and down sides, i.e., $\phi_{u}$ and $\phi_{d}$ versus $\theta_{in}$, is plotted as black and blue lines in the second row. The GH shift given by the stationary-phase method of the outgoing beam to the up and down side, i.e., $S_{GH,u}$ and $S_{GH,d}$ versus $\theta_{in}$, are plotted as black and blue lines in the third row. In the first and second column, the incident angle is positive; in the third and fourth column, the incident angle is negative. In the first and third column, $\varphi=\pi/2$; in the second and fourth column, $\varphi=3\pi/2$. The frequency of the TM polarized incident plane wave is $af/c=0.380055$. The structural parameters of the meta-grating are the same as those in Fig. \ref{figure_bandUGR}(b-f). The vertical dashed line marks the resonant angle at $\pm21.5^{o}$. }
\label{figure_grateGH}
\end{figure*}

As the frequency and lateral component wave number are resonant with the frequency and $k_{x}$ of a quasi-BIC, the GH shift could be highly enhanced. For the scheme in Fig. \ref{figure_scheme} that inducing coherent GH shift, the far-field pattern under plane wave incidence can be given as
\begin{equation}
E_{y}=\{\begin{array}{cc}
c_{in}e^{i\varphi}e^{ik_{x}x-ik_{z}}+c_{up}e^{ik_{x}x+ik_{z}} & z\rightarrow+\infty  \\
c_{in}e^{ik_{x}x+ik_{z}}+c_{down}e^{ik_{x}x-ik_{z}} & z\rightarrow-\infty \\
\end{array}
\end{equation}
, where $k_{x}=n_{b}k_{0}\sin\theta_{in}$, $k_{z}=\sqrt{n_{b}^{2}k_{0}-k_{x}^{2}}$, $k_{0}=2\pi /\lambda$ with $\lambda$ and $\theta_{in}$ being the wavelength and incident angle of the incident plane wave. In the numerical calculation, $c_{in}$ is set to be unity, and $c_{up(down)}$ are extracted from the near field pattern of $E_{y}$ by performing Fourier transform on the up and down boundary of the computational domain. In the FEM calculation, total-field scatter-field interfaces are added to the physical domains above and below the meta-grating to simulate the incident plane wave from up and down side, respectively \cite{theFEM1,theFEM3}. The total incident power is $P_{inc}\propto2|c_{in}|^{2}$. The outgoing powers to the up and down sides are $P_{u}\propto|c_{up}|^{2}$ and $P_{d}\propto|c_{down}|^{2}$, respectively. The phase of the outgoing plane wave to the up and down sides is $\phi_{u}=arg(c_{up})$ and $\phi_{d}=arg(c_{down})$, respectively. According to the stationary-phase method, the GH shifts of the outgoing beams to the up and down sides are given as
\begin{equation}
S_{GH,u(d)}=-\frac{\lambda}{2\pi}\frac{\partial\phi_{u(d)}}{\theta_{in}}
\end{equation}
As $\varphi$ changes, the near field pattern $E_{y}$ changes, which in turn changes $c_{up,down}$ and $\phi_{u(d)}$. Thus, $S_{GH,u(d)}$ is dependent on $\varphi$.

\begin{figure}[tbp]
\scalebox{0.6}{\includegraphics{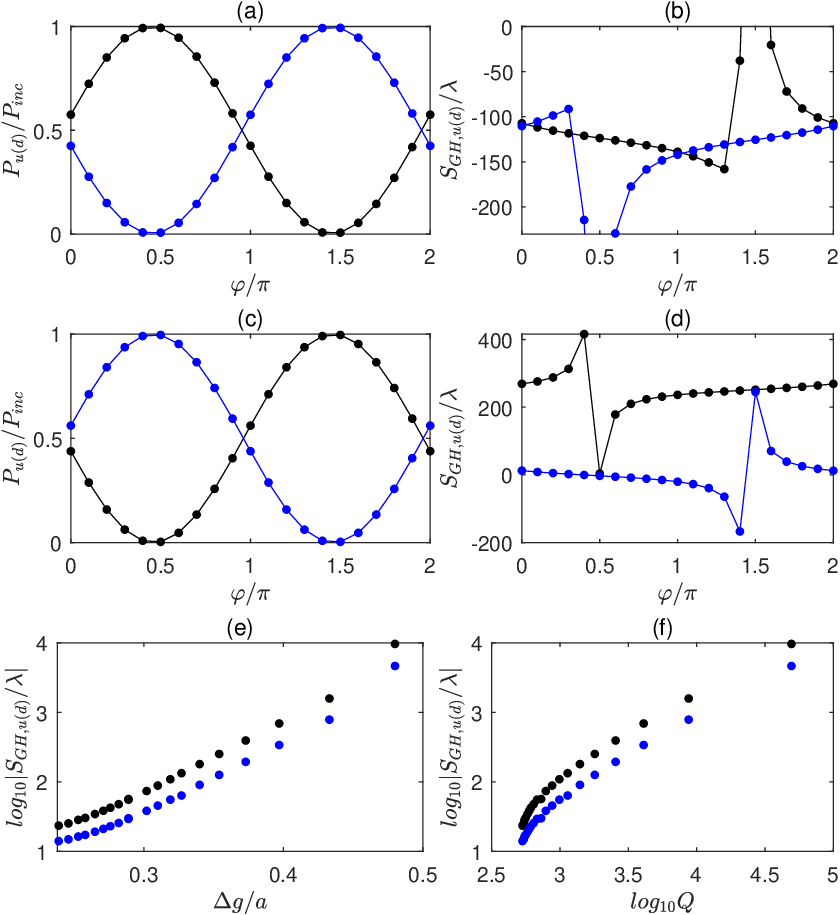}}
\caption{ For the system in Fig. \ref{figure_grateGH}, as the incident angle is fixed to the resonant angle at $\pm29.5^{o}$, $P_{u}/P_{inc}$ and $P_{d}/P_{inc}$ versus $\varphi$ are plotted as black and blue dots, respectively, in panels (a,c); $S_{GH,u}$ and $S_{GH,d}$ versus $\varphi$ are plotted as black and blue dots in panels (b,d). For the case with $+21.5^{o}$ and $-21.5^{o}$, the results are plotted in (a,b) and (c,d), respectively. (e) For the branches of systems in Fig. \ref{figure_paraUGR}, as $\varphi=3\pi/2$, the magnitude of the GH shift of the outgoing wave to the down and up sides, with $\theta_{in}=+\sin^{-1}[|k_{x}|/(n_{b}k_{0})]$ and $\theta_{in}=-\sin^{-1}[|k_{x}|/(n_{b}k_{0})]$, versus $\Delta p$, are plotted as blue and black dots, respectively. (f) The same as (e), but substitute the x axis by the corresponding Q factor of the resonant modes.  }
\label{figure_GHphase}
\end{figure}

For the system in Fig. \ref{figure_bandUGR}, the numerical results of the GH shift given by the stationary-phase method are plotted in Fig. \ref{figure_grateGH}. The wavelength of the incident plane wave is resonant with the resonant frequency of the UGR and CPS, i.e., $\lambda=c/f$. The incident angle is scaled across the resonant angle given as $\sin^{-1}[k_{x}/(n_{b}k_{0})]$, with $k_{x}$ being the wave number of the UGR(CPS), and $k_{0}=2\pi f/c$. As $\theta_{in}>0$, the CPS with $k_{x}>0$ is resonantly excited. As $\varphi=\pi/2$ and $3\pi/2$, $P_{u}$ and $P_{d}$ have a resonant peak near to the resonant angle, while $P_{d}$ and $P_{u}$ have a resonant dip near to the resonant angle, as shown in Fig. \ref{figure_grateGH}(a) and (b), respectively. Thus, at the resonant angle, $\varphi$ controls the direction of the outgoing power. When $\varphi$ is equal to $\pi/2$ and $3\pi/2$, the outgoing power exits to the up and down sides, respectively. Within the peak and dip, the phase of the outgoing plane wave $\phi_{u(d)}$ is highly sensitive to $\theta_{in}$, as shown in Fig. \ref{figure_grateGH}(e,f). Thus, the GH shifts of the outgoing beams have large magnitude, as shown in Fig. \ref{figure_grateGH}(i,j). At the resonant angle, as $\varphi$ is equal to $\pi/2$ and $3\pi/2$, $|S_{GH,d}|$ and $|S_{GH,u}|$ become ultra-large, but the corresponding outgoing power is near to zero, i.e., $P_{d}$ and $P_{u}$ are near to zero, respectively. For the incidence of Gaussian beams, the GH shifts are averaged of $S_{GH,d(u)}$ within the dip of $P_{d(u)}$, so that the numerical value of the GH shifts are not ultra-large. On the other hand, as $\varphi$ is equal to $\pi/2$ and $3\pi/2$, $S_{GH,u}$ and $S_{GH,d}$ are equal to $-123.8\lambda$, and $P_{u}$ and $P_{d}$ are near to $P_{inc}$. Thus, as $\varphi$ is equal to $\pi/2$ and $3\pi/2$, the majority of outgoing power goes into the outgoing beam to the up and down sides, respectively, with sizable negative GH shift. As $\varphi$ varies from $0$ to $2\pi$, $P_{u}$ and $P_{d}$ at the resonant angle oscillate with opposite trend, as shown by the black and blue dots in Fig. \ref{figure_GHphase}(a). The corresponding GH shifts of the outgoing beams to the up and down sides, i.e., $S_{GH,u}$ and $S_{GH,d}$ versus $\varphi$ are plotted as black and blue dots in Fig. \ref{figure_GHphase}(b). As $\varphi$ varies near  $\pi/2$, $P_{u}$ is approximately equal to $P_{inc}$, while $S_{GH,u}$ changes with the sensitivity to $\varphi$ being given as $\frac{\partial S_{GH,u}}{\partial\varphi}\approx(-0.1406\lambda)$ $deg^{-1}$. On the other hand, as $\varphi$ varies near  $3\pi/2$, the sensitivity of $S_{GH,d}$ to $\varphi$ is $(+0.1406\lambda)$ $deg^{-1}$.

As $\theta_{in}<0$, the UGR with $k_{x}<0$ is resonantly excited. As $\varphi=\pi/2$ and $3\pi/2$, $P_{d}$ and $P_{u}$ are equal to $P_{inc}$, and are hardly dependent on $\theta_{in}$, as shown in Fig. \ref{figure_grateGH}(c) and (d), respectively. Thus, incident Gaussian beams with sizable divergence angle could be completely scattered to the down and up sides, as $\varphi=\pi/2$ and $3\pi/2$, respectively. As $\varphi=\pi/2$, $\phi_{u(d)}$ are slightly dependent on $\theta_{in}$, as shown in Fig. \ref{figure_grateGH}(g). Thus, the corresponding GH shifts have small magnitude, as shown in Fig. \ref{figure_grateGH}(k). As $\varphi=3\pi/2$, $\phi_{u(d)}$ are highly sensitive to $\theta_{in}$ near to the resonant angle, as shown in Fig. \ref{figure_grateGH}(h). Thus, the corresponding GH shifts have sizable magnitude, as shown in Fig. \ref{figure_grateGH}(l). As $\varphi$ varies from $0$ to $2\pi$, $P_{u}$ and $P_{d}$ at the resonant angle oscillate with opposite trend, as shown by the black and blue dots in Fig. \ref{figure_GHphase}(c). The corresponding GH shift of the outgoing beams to the up and down side, i.e., $S_{GH,u}$ and $S_{GH,d}$ versus $\varphi$ are plotted as black and blue dots in Fig. \ref{figure_GHphase}(d). As $\varphi$ varies near  $3\pi/2$, $P_{u}$ is approximately equal to $P_{inc}$, so that the majority of outgoing energy goes to the outgoing beam to the up side. The sensitivity of $S_{GH,u}$ to $\varphi$ is $(+0.1426\lambda)$ $deg^{-1}$.

To sum up, in three specific cases with $|\theta_{in}|$ being equal to the resonant angle of the UGR(CPS), the majority of outgoing energy goes to one side of the grating, and has sizable GH shift: (i) when $\theta_{in}>0$ and $\varphi=\pi/2$, $P_{u}\approx P_{inc}$ and $S_{GH,u}$ is negative and sizable; (ii) when $\theta_{in}>0$ and $\varphi=3\pi/2$, $P_{d}\approx P_{inc}$ and $S_{GH,d}$ is negative and sizable; (iii) when $\theta_{in}<0$ and $\varphi=3\pi/2$, $P_{u}\approx P_{inc}$ and $S_{GH,u}$ is positive and sizable. The cases (i) and (ii) are symmetric to each other with the same value of GH shift. As the structural parameters change within the subset of systems with UGR and CPS, the magnitudes of the GH shifts of the three cases change. For the series of systems in Fig. \ref{figure_paraUGR}, the magnitude of GH shift of case (ii) and (iii) versus $\Delta g$ are plotted as blue and black dots in Fig. \ref{figure_GHphase}(e), respectively. As $\Delta g$ approaches $0.5$, the Q factor of the UGR(CPS) in the corresponding system exponentially increases, so that the magnitudes of the GH shifts also increase. By plotting the magnitudes of the GH shifts versus the Q factor in Fig. \ref{figure_GHphase}(f), one can find that the magnitudes of the GH shifts are approximately proportional to the Q factor.

Interestingly, the large GH shift of excitation of the CPS is accompanied by a resonant peak of reflectance (transmittance), while that of the UGR is accompanied by a constant reflectance (transmittance). The phenomenon can be explained by applying temporal coupled mode theory (TCMT) \cite{tcmt01,tcmt02,tcmt03}. For a given $|k_{x}|$, we denote the amplitude of the modes at the band structure with $k_{x}=|k_{x}|$ and $k_{x}=-|k_{x}|$ as $a_{1}$ and $a_{2}$, respectively. The resonant frequency and decay rate of the two modes are designated as $\omega_{b}$ and $\gamma$, which are equal to real and imaginary part of the eigen frequency. When $|ak_{x}/2\pi|=0.201$, the two modes with amplitude being $a_{1}$ and $a_{2}$ are CPS and UGR in Fig. \ref{figure_bandUGR}, respectively. $k_{x}$ and $\omega_{b}$ of the two topological modes are designated as $k_{x,r}$ and $\omega_{r}$, respectively. $\gamma$ is approximated as constant. For a given $|k_{x}|$, the four ports of plane waves are indexed by $j\in[1,2,3,4]$, which are the ports to the upper-left, lower-left, upper-right, lower-right of the resonant modes. The amplitude of the incident fields from the ports and the outgoing fields to the ports are designated as $p_{j}$ and $u_{j}$, respectively. Incident fields from $p_{1,2}$ ($p_{3,4}$ )can only excite $a_{1}$ ($a_{2}$), and generate outgoing fields $u_{3,4}$ ($u_{1,2}$). The dynamic equations of the modes are given as
\begin{equation}
\frac{d}{dt}a_{1}=(-i\omega_{b}-\gamma)a_{1}+\kappa_{1}p_{1}+\kappa_{2}p_{2}
\end{equation}
and
\begin{equation}
\frac{d}{dt}a_{2}=(-i\omega_{b}-\gamma)a_{2}+\kappa_{3}p_{3}+\kappa_{4}p_{4}
\end{equation}
where $\kappa_{j}$ are the coupling coefficient from incident field $p_{j}$ to the corresponding resonant mode. The outgoing field in each port is generated by direct scattering of the incident field and radiation from the resonant mode. The direct scattering can be approximated as reflection and transmission of a uniform dielectric slab with thickness being $2t+G$. Thus, the reflection and transmission coefficient of the incident field from a single port are approximately equal to $-1/\sqrt{2}$ and $i/\sqrt{2}$. For the radiation from the resonant modes, we consider the cases with $k_{x}=k_{x,r}$, i.e., excitation of CPS or UGR. The radiation rate of the CPS $a_{1}$ to ports $u_{3}$ and $u_{4}$ are $d_{3}=\sqrt{\gamma}$ and $d_{4}=-i\sqrt{\gamma}$, respectively, which satisfy the condition of energy preservation $2\gamma=|d_{3}|^{2}+|d_{4}|^{2}$, and the condition of the CPS $d_{3}/d_{4}=-i$. Similarly, the radiation rate of the UGR $a_{2}$ to ports $u_{1}$ and $u_{2}$ are $d_{1}=\sqrt{2\gamma}$ and $d_{2}=0$, respectively. By applying the condition of time-reversal symmetry, one can find that $\kappa_{1}=d_{1}=\sqrt{2\gamma}$, $\kappa_{2}=d_{2}=0$, $\kappa_{3}=d_{3}=\sqrt{\gamma}$, and $\kappa_{4}=d_{4}=-i\sqrt{\gamma}$. For the other resonant modes with $k_{x}\approx k_{x,r}$, the coefficient $d_{j}$ and $\kappa_{j}$ are assumed to be constant. When the frequency of the incident fields is fixed to be $\omega$, the excitation of the resonant modes are given as
\begin{equation}
a_{1}=\frac{\sqrt{2\gamma}p_{1}}{-i(\omega-\omega_{b})+\gamma}
\end{equation}
and
\begin{equation}
a_{2}=\frac{\sqrt{\gamma}p_{3}-i\sqrt{\gamma}p_{4}}{-i(\omega-\omega_{b})+\gamma}
\end{equation}
Summing up the direct scattering of the incident field and the radiation from the resonant mode, the outgoing field are given as
\begin{eqnarray}
u_{1}=&&\frac{1}{\sqrt{2}}(-1+\frac{2\gamma}{-i(\omega-\omega_{b})+\gamma})p_{3}\nonumber\\&&+\frac{i}{\sqrt{2}}(1-\frac{2\gamma}{-i(\omega-\omega_{b})+\gamma})p_{4}
\end{eqnarray}
\begin{equation}
u_{2}=\frac{i}{\sqrt{2}}p_{3}-\frac{1}{\sqrt{2}}p_{4}
\end{equation}
\begin{equation}
u_{3}=\frac{1}{\sqrt{2}}(-1+\frac{2\gamma}{-i(\omega-\omega_{b})+\gamma})p_{1}+\frac{i}{\sqrt{2}}p_{2}
\end{equation}
\begin{equation}
u_{4}=\frac{i}{\sqrt{2}}(1-\frac{2\gamma}{-i(\omega-\omega_{b})+\gamma})p_{1}-\frac{1}{\sqrt{2}}p_{2}
\end{equation}
In the following, we consider the case with $\omega=\omega_{r}$ and varying $k_{x}$.

We firstly consider the case with incidence field from the left ports, i.e., $p_{1}=1$ and $p_{2}=\pm i$. With $p_{2}=-i$, the amplitude of outgoing fields are  $u_{3}=\sqrt{2}\gamma/[-i(\omega_{r}-\omega_{b})+\gamma]$, and $u_{4}=\sqrt{2}(\omega_{r}-\omega_{b})/[-i(\omega_{r}-\omega_{b})+\gamma]$. In the angular spectrum, $\theta_{in}$ is scanned, so that the lateral wave number $k_{x}=(n_{b}\omega_{r}/c)\sin\theta_{in}$ is scanned, which in turn scans $\omega_{b}$. The band structure near the resonant mode at $(k_{x,r},\omega_{r})$ can be approximated as $\omega_{b}(k_{x})=\omega_{r}+\omega_{r,1}(k_{x}-k_{x,r})$, with $\omega_{r,1}$ being the slope of the band structure at $(k_{x,r},\omega_{r})$. Thus, $u_{3}$ and $u_{4}$ are functions of $\theta_{in}$ with typical Lorentzian profile of resonant peak and dip, respectively. They are given as
\begin{equation}
u_{3}=\frac{\sqrt{2}\gamma}{\gamma+i\omega_{r,1}[(n_{b}\omega_{r}/c)\sin\theta_{in}-k_{x,r}]}
\end{equation}
and
\begin{equation}
u_{4}=\frac{-\sqrt{2}\omega_{r,1}[(n_{b}\omega_{r}/c)\sin\theta_{in}-k_{x,r}]}{\gamma+i\omega_{r,1}[(n_{b}\omega_{r}/c)\sin\theta_{in}-k_{x,r}]}
\end{equation}
with peak and dip at $\theta_{in}=sin^{-1}[ck_{x,r}/(n_{b}\omega_{r}]$. Absolution value and complex angle of $u_{3,4}$ give the line shapes that can approximate the numerical value in Fig. \ref{figure_grateGH}(a) and (e), respectively. Both reflectance and GH shift have resonant peak at the resonant angle. For the case with $p_{2}=i$, the amplitude of outgoing fields are  $u_{3}=-i\sqrt{2}(\omega_{r}-\omega_{b})/[-i(\omega_{r}-\omega_{b})+\gamma]$, and $u_{4}=-i\sqrt{2}\gamma/[-i(\omega_{r}-\omega_{b})+\gamma]$, which have typical Lorentzian profile with resonant dip and peak, respectively. In these two cases, the direct scattering of the two incident fields have destructive interference with each other, so that the outgoing fields are purely from the radiation of the CPS, which have typical Lorentzian profile.

Secondly, we consider the case with incidence field from the right ports, i.e., $p_{3}=1$ and $p_{4}=\pm i$. When $p_{4}=-i$, the amplitude of outgoing fields are $u_{1}=0$ and $u_{2}=i\sqrt{2}$. Both absolute value and complex angle of $u_{1,2}$ are constant, which approximately match with the numerical result in Fig. \ref{figure_grateGH}(c) and (g). When $p_{4}=i$, the amplitude of outgoing fields are $u_{1}=\sqrt{2}[i(\omega_{r}-\omega_{b})+\gamma]/[-i(\omega_{r}-\omega_{b})+\gamma]$ and $u_{2}=0$. The absolute value of $u_{1}$ is constantly equal to unity, but the complex angle of $u_{1}$ have a resonant peak, which match with the numerical result in Fig. \ref{figure_grateGH}(d) and (h). The complex angle of $u_{1}$ is given as
\begin{equation}
Arg(u_{1})=2tan^{-1}\frac{-\omega_{r,1}[(n_{b}\omega_{r}/c)\sin\theta_{in}-k_{x,r}]}{\gamma}
\end{equation}
In these two cases, both incident fields from up and down sides can excite the UGR. When $p_{4}=-i$, the excitation of the UGR by the incident fields from up and down sides have destructive interference with each other, so that the UGR is not excited. The outgoing fields are purely from the direct scattering, which have not GH shift. When $p_{4}=i$, the UGR is excited, and the direct scattering of the two incident fields have constructive interference with the radiation from the UGR to port one. The constructive interference flatten the magnitude of $u_{1}$, but keep the fast variation of  $Arg(u_{1})$, so that the enhancement of the GH shift is accompanied with a constant output energy spectrum.

Finally, we consider the case with  $k_{x}=k_{x,r}$, $\omega_{b}=\omega_{r}$, and varying $\varphi$, i.e., $p_{1}=p_{3}=1$, $p_{2}=p_{4}=e^{i\varphi}$. In this case, the solution of the outgoing fields are simplified as $u_{1}=\frac{1}{\sqrt{2}}(1-ie^{i\varphi})$, $u_{2}=\frac{1}{\sqrt{2}}(i-e^{i\varphi})$, $u_{3}=\frac{1}{\sqrt{2}}(1+ie^{i\varphi})$, and $u_{4}=\frac{1}{\sqrt{2}}(-i-e^{i\varphi})$. The absolute value of $u_{j}$ match with the numerical results in Fig. \ref{figure_GHphase}(a) and (c).

\subsection{Gaussian beam incidence}

\begin{figure}[tbp]
\scalebox{0.59}{\includegraphics{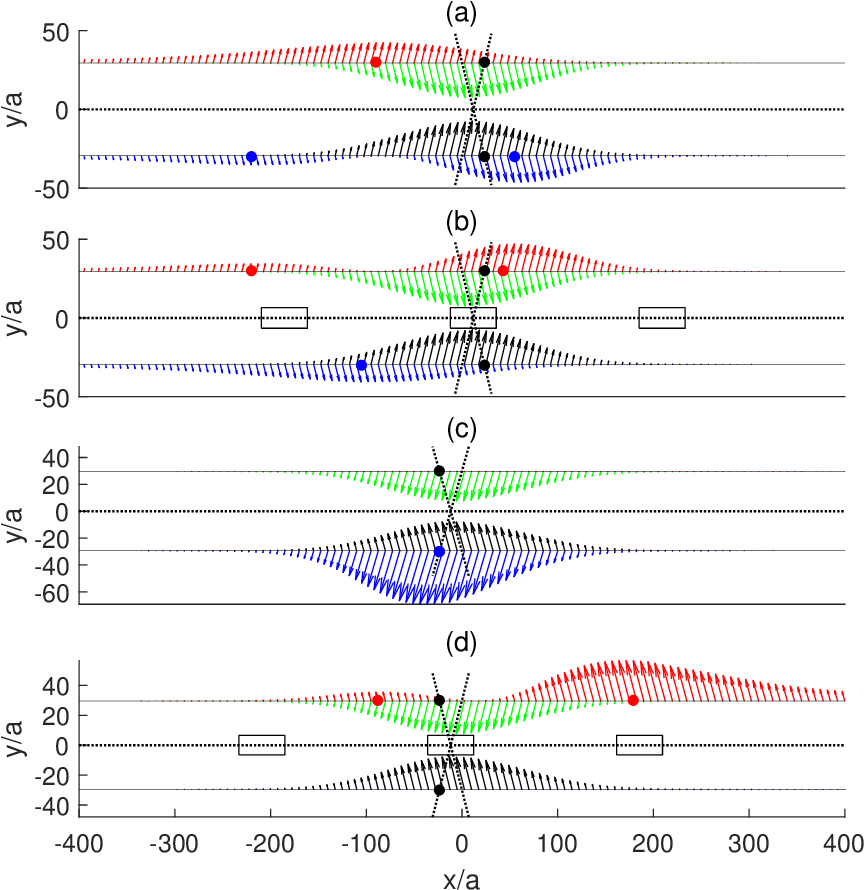}}
\caption{ For a Gaussian beam with $af/c=0.380055$, $\theta_{Inc}=21.5^{o}$, and $w_{0}=50\lambda$, the spatial distribution of time-averaged Poynting vector of the incident and the outgoing beams in the up side of the grating along the horizontal observation plane at $z=30a$ are plotted as green and red arrows, respectively; those at $z=-30a$ are plotted as black and blue arrows, respectively. The axes of the incident and regular outgoing beams without GH shift are plotted as black dashed line. The black dots mark the axes of the regular outgoing beams at $z=\pm30a$. The meta-grating locates at $z=0$. The red and blue dots mark the axes of the outgoing beams at $z=30a$ and $-30a$, respectively. The structural parameters of the meta-grating are the same as those in Fig. \ref{figure_bandUGR}(b-f).  }
\label{figure_slabGH}
\end{figure}

\begin{figure}[tbp]
\scalebox{0.42}{\includegraphics{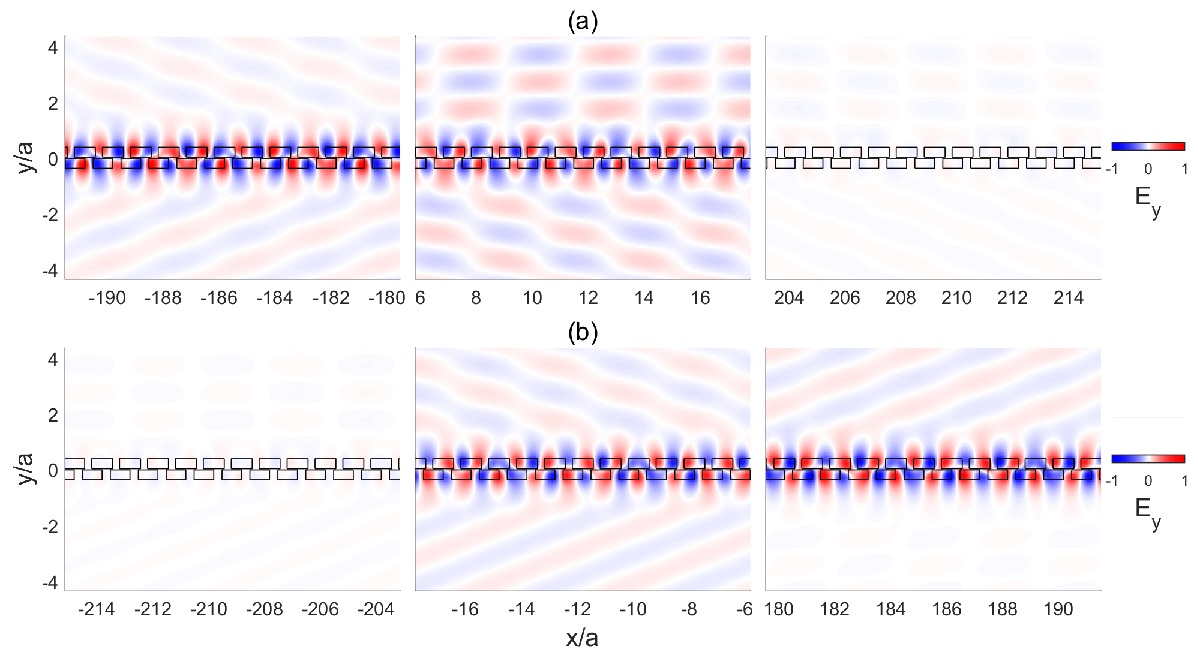}}
\caption{ The field pattern of $E_{y}$ within the rectangularly surrounded regions in Fig. \ref{figure_slabGH}(b) and (d) are plotted in panel (a) and (b), respectively.  }
\label{figure_slabGHfield}
\end{figure}

In this subsection, the coherent GH shift of the outgoing beams under incidence of two Gaussian beams with finite beam width is studied. As the beam width approaches infinity, the Gaussian beams are equivalent to plane waves, so that the results are the same as those in the previous subsection. As the beam width becomes finite, a Gaussian beam can be considered as a superposition of multiple plane waves with incident angular spectrum being a Gaussian function. Thus, the GH shift of the scattering field can be theoretically obtained by convolution between the incident angular spectrum and the angularly dependent GH shift given by the stationary-phase method. In this paper, the coherent GH shift with incidence of Gaussian beams is studied by numerically calculating the near field pattern with FEM.

We consider the incident Gaussian beams with finite beam width in the x-z plane, but infinite beam width along the y axis (i.e. the field is uniform along y direction). For the Gaussian beams from up and down sides with fixed wavelength and beam width, the field profile in the x-z plane is given as
\begin{equation}
\mathbf{E}_{\kappa}(\rho,\eta)=\hat{y}E_{y,\kappa}\sqrt{\frac{w_{0}}{w(\xi)}}e^{-\frac{\rho^{2}}{w^{2}(\xi)}+ik_{0}\xi+ik_{0}\frac{\rho^{2}}{2R(\xi)}-\frac{i}{2}\eta(\xi)} \label{gaussianBeam}
\end{equation}
, where $\kappa=\pm1$ represents incident Gaussian beams from up and down sides,  $\xi=(\mathbf{r}-\mathbf{r}_{p0})\cdot\mathbf{k}_{Inc,\kappa}$, $\rho=|\mathbf{r}-\mathbf{r}_{p0}-\xi\mathbf{k}_{Inc,\kappa}|$, $w(\xi)=w_{0}\sqrt{1+(\xi/z_{0})^{2}}$, $R(\xi)=\xi[1+(z_{0}/\xi)^{2}]$, $\eta(\xi)=\tan^{-1}(\xi/z_{0})$, $z_{0}=\frac{k_{0}w_{0}^{2}}{2}$, with $w_{0}$ being the beam width at the beam waist, $\mathbf{r}_{p0}$ being the location of the focus point, $\mathbf{k}_{Inc}=-\kappa\cos\theta_{Inc}\hat{z}+\sin\theta_{Inc}\hat{x}$ being the unit vector along the incident direction $\theta_{Inc}$ from the normal line. The amplitude of the two Gaussian beams is $E_{y,\kappa}$ with $E_{y,+1}/E_{y,-1}=e^{i\varphi}$. The focus of the beam is at the center of the meta-grating. Performing Fourier transformation to the field profile near $\mathbf{r}_{p0}$, the Gaussian beam can be expanded into a superposition of a serial of plane waves with the same wavelength and varying incident angle. The incident angular spectrum is given by Gaussian function $\Theta(\theta)=e^{-[(\theta-\theta_{Inc})/\Delta\theta_{in}]^{2}/2}$, with $\Delta\theta_{in}=\sin^{-1}[\lambda/(\sqrt{2}\pi w_{0})]$ being the divergence angle. If the divergence angle is much smaller than the width of the resonant peak in Fig. \ref{figure_grateGH}(i-l), the Gaussian beams can be considered as plane waves.

For the system in Fig. \ref{figure_bandUGR}, the numerical results of the GH shift under incidence of the Gaussian beam with wavelength being $\lambda=a/0.380055$, $\theta_{Inc}=\pm21.5^{o}$, and $w_{0}=50\lambda$ are plotted in Fig. \ref{figure_slabGH}. The divergence angle is equal to $0.258^{o}$, which is larger than the width of the resonant peak in Fig. \ref{figure_grateGH}(i-l), so that the outgoing beams to each side of the grating is usually consisted of two parallel beams with opposite GH shifts. One of the beams is due to the GH shift at the resonant angle, while the other beam is due to the GH shift outside of the resonant peak. Four specific cases are studied:

(i) $\theta_{Inc}=+21.5^{o}$ and $\varphi=\pi/2$. The outgoing beam to the up side has GH shift to the left. The axis of the beam, that is approximated as the location with maximum magnitude of the Poynting vector at $y=30a$, is marked by the red dot in Fig. \ref{figure_slabGH}(a). The GH shift can be approximated by the distance between the red dot and the black dot at $y=30a$, which is $-41.3\lambda$. The magnitude is smaller than that given by the stationary-phase method at the resonant angle [ $S_{GH,u}=-123.8\lambda$ as shown in Fig. \ref{figure_grateGH}(i)]. As $w_{0}$ becomes larger, the magnitude of the GH shift with Gaussian beams' incidence will approaches the theoretical value given by the stationary-phase method. The outgoing beam to the down side is due to the angular spectrum both inside and outside of the dip of $P_{d}$, which has negative and positive GH shift. Thus, the outgoing beam consists of two parallel beams with opposite GH shifts. The numerical value of the GH shifts are $10.8\lambda$ and $-89.9\lambda$.

(ii) $\theta_{Inc}=+21.5^{o}$ and $\varphi=3\pi/2$. The outgoing beams are approximately equal to the upside down of those in the previous case with $\theta_{Inc}=+21.5^{o}$ and $\varphi=\pi/2$, as shown in Fig. \ref{figure_slabGH}(b). The field pattern of $E_{y}$ within the three rectangular regions is plotted in Fig. \ref{figure_slabGHfield}(a). In the middle region, the two incident beams interfere with each other, and then couple the optical field into the meta-grating. The CPS shown in Fig. \ref{figure_bandUGR}(f) is excited. At the green dot of Fig. \ref{figure_bandUGR}(b), the slope of the band is negative, so that the group velocity of the CPS is negative. Thus, the optical field within the meta-grating travels toward the left direction, and reaches the left rectangular region in Fig. \ref{figure_slabGH}(b), as shown by the strong field pattern in Fig. \ref{figure_slabGHfield}(a). Meanwhile, the left traveling CPS radiates energy to both up and down sides of the grating, which form the outgoing beams with GH shift to the left. The right rectangular region in Fig. \ref{figure_slabGH}(b) is outside of the beam width of the incident Gaussian beams, so that the optical field within the meta-grating is weak, as shown in Fig. \ref{figure_slabGHfield}(a).

(iii) $\theta_{Inc}=-21.5^{o}$ and $\varphi=\pi/2$. The incident Gaussian beams do not excite the UGR, so that the outgoing beams do not have GH shift, as shown in Fig. \ref{figure_slabGH}(c).

(iv) $\theta_{Inc}=-21.5^{o}$ and $\varphi=3\pi/2$. The incident Gaussian beam excites the UGR, so that the outgoing beams have GH shifts. Because of the unidirectional radiation from the UGR, all outgoing energy goes to the outgoing beam to the up side. The outgoing beam to the up side is split into two parallel beams with minority and majority of the outgoing power, whose numerical GH shifts are $-22.2\lambda$ and $77.4\lambda$. Because the Gaussian beams are incident from the right side of the normal line, a positive value of the GH shift given by the stationary-phase method (GH shift to the right) represents negative GH shift relative to the incident beams, and vice versa. Because the group velocity of the UGR is positive, after the incident Gaussian beams couple the optical field into the meta-grating, the optical field travels to the right, and reaches the right rectangular region, as shown in Fig. \ref{figure_slabGHfield}(b). The unidirectional radiation from the UGR to the up side of the grating can be visualized by the plane wave in the right rectangular region in Fig. \ref{figure_slabGHfield}(b).

More numerical results of Gaussian beams' incidence with varying $\varphi$, and the tuning of the GH shift by $\varphi$, are given in the appendix. The maximum magnitude of the sensitivity of the GH shift to $\varphi$ is $0.2111\lambda$ $deg^{-1}$. If the phase delay device in Fig. \ref{figure_scheme} contain an air chamber with length being $100\lambda$, as the refractive index of air change from unity to $1.0001$, the change of $\varphi$ is $3.6^{o}$, so that the change of  $S_{GH}$ is $0.76\lambda$, which can be measured by optical imaging of the outgoing beam. Thus, the system can be applied as refractive index sensing device of gas.

\section{Conclusion}

In conclusion, the scheme to generate a coherent GH shift by tuning the relative phase of the two incident coherent optical beams is proposed. The two incident optical beams have the same frequency and $k_{x}$, but opposite $k_{z}$. As the frequency and $k_{x}$ of the incident beams are resonant with those of a topological resonant mode, such as quasi-BIC, UGR, or CPS, the coherent GH shift is enhanced. The branches of UGRs and CPSs are subset of the quasi-BICs, which are obtained by simultaneously tuning multiple parameters from BIC. As the UGRs and CPSs approach the BIC in the parameter space, the Q factor exponentially increases, which induces Goos-H$\ddot{a}$nchen shifts with large magnitude. The value of the coherent GH shift is sensitive to the phase difference between the two incident optical beams. As the incident beams resonantly excite a CPS, the direct scattering of the two incident beams cancel each other, so that the spectrum of the outgoing field is purely due to the radiation from the CPS. On the other hand, as the incident beams resonantly excite a UGR, the direct scattering of the two incident beams have constructive interference with the radiation from the UGR. The different process of interference results in different behavior of the enhancement of the GH shift. As the CPS or UGR is excited, the peak of the GH shift is accompanied with angular spectrum of the outgoing energy flux with a peak or constant profile, respectively. As the incident fields being Gaussian beams with finite beam width, the resonant excitation of CPS or UGR induces localized energy flux inside the meta-grating. The energy flux travels a distance, which is approximately equation to the GH shift given by the stationary-phase method, and keep radiating energy to the background, which form the outgoing beams with GH shift. The sensitivity of the GH shift to $\varphi$ can be applied in refractive index sensing devices.

\begin{acknowledgments}
This project is supported by the Natural Science Foundation of Guangdong Province of China (Grant No. 2022A1515011578),  the Special Projects in Key Fields of Ordinary Universities in Guangdong Province(New Generation Information Technology, Grant No. 2023ZDZX1007), and the Startup Grant at Guangdong Polytechnic Normal University (Grant No. 2021SDKYA117).
\end{acknowledgments}

\section{appendix}

\begin{figure*}[tbp]
\scalebox{0.53}{\includegraphics{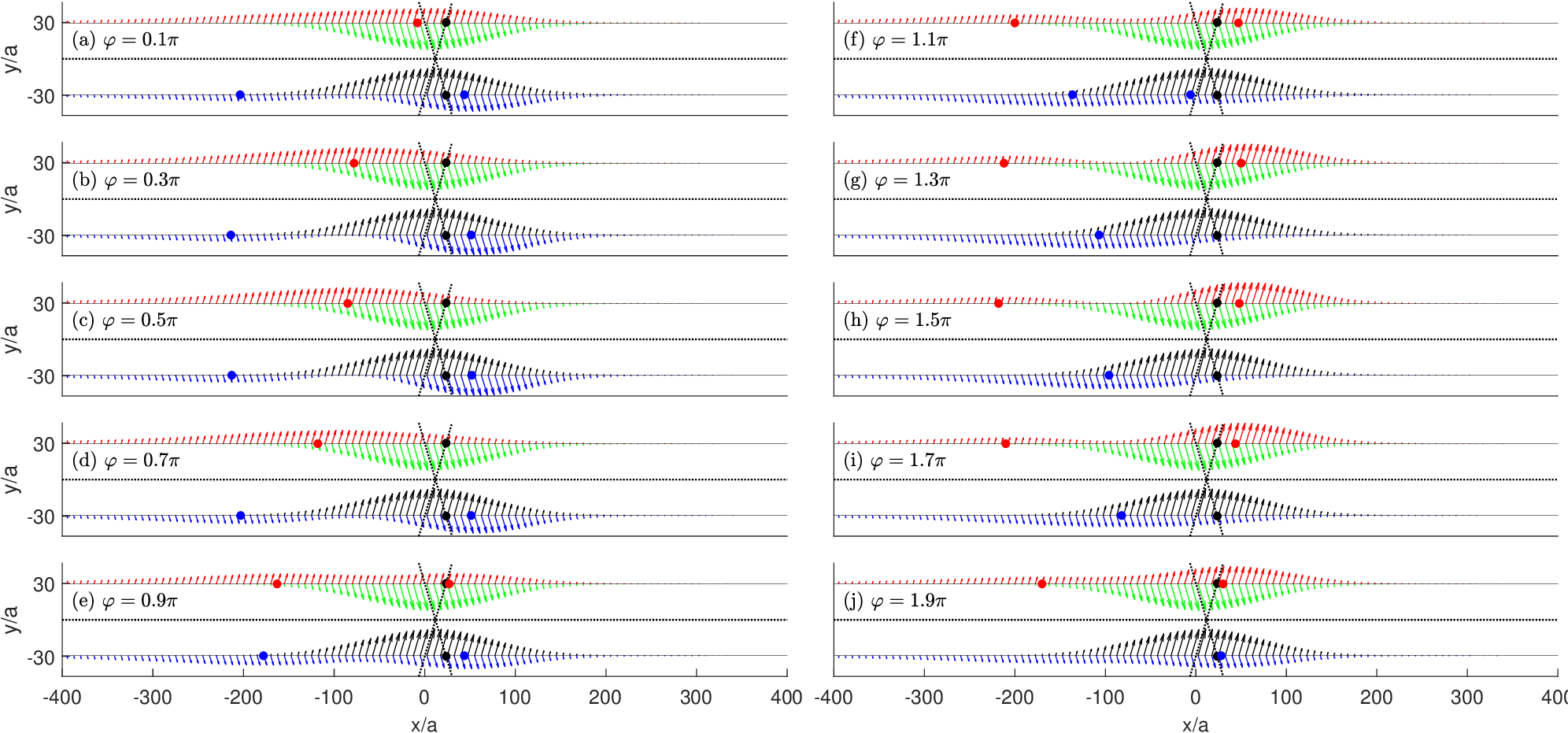}}
\caption{ The same as Fig. \ref{figure_slabGH}(a,b), with $\theta_{Inc}=21.5^{o}$ and $\varphi$ varying from $0.1\pi$ to $1.9\pi$ with interval of $0.2\pi$ in panel (a) to (j). }
\label{figure_slabGHt+all}
\end{figure*}

\begin{figure*}[tbp]
\scalebox{0.53}{\includegraphics{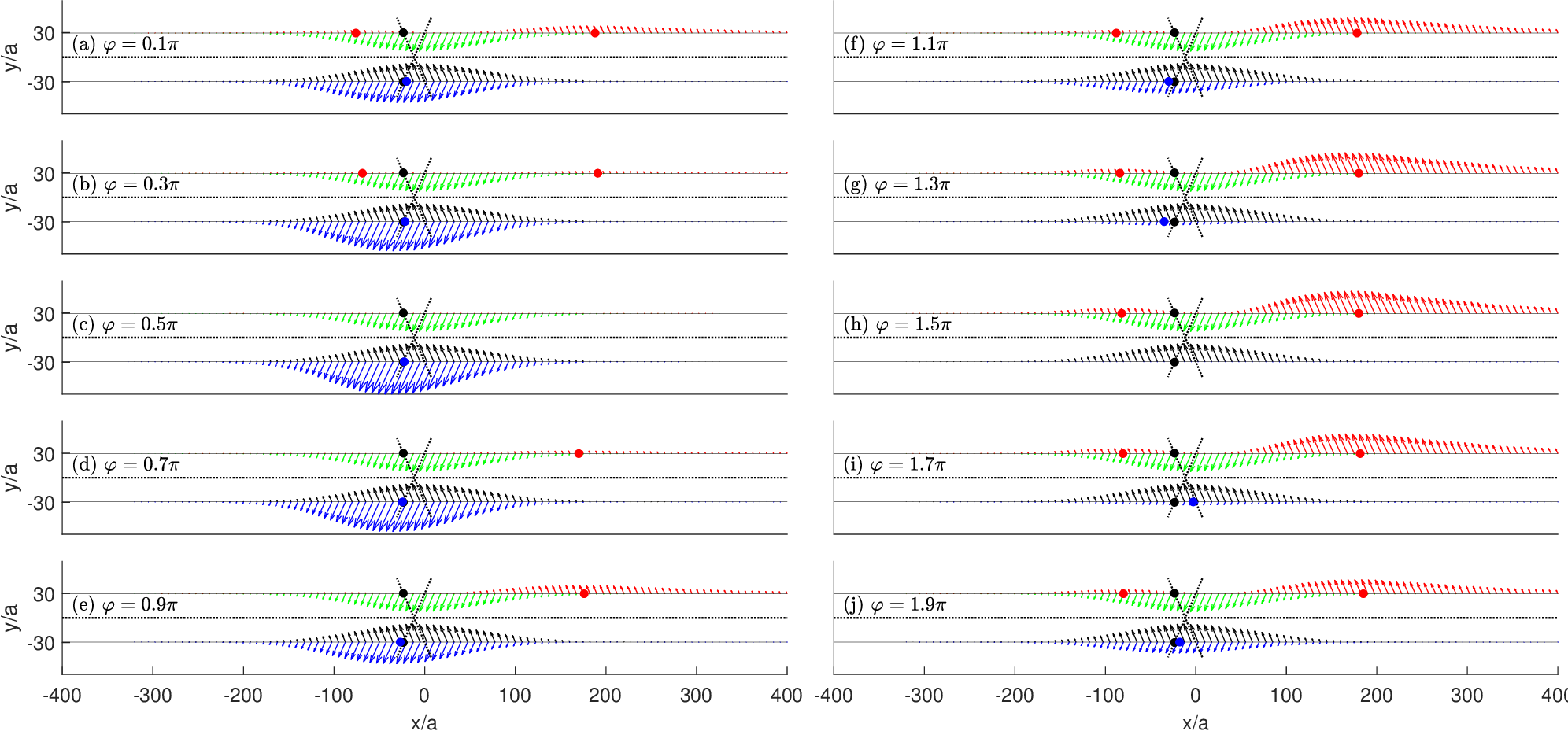}}
\caption{ The same as Fig. \ref{figure_slabGH}(c,d), with $\theta_{Inc}=-21.5^{o}$ and $\varphi$ varying from $0.1\pi$ to $1.9\pi$ with interval of $0.2\pi$ in panel (a) to (j). }
\label{figure_slabGHt-all}
\end{figure*}

\begin{figure}[tbp]
\scalebox{0.6}{\includegraphics{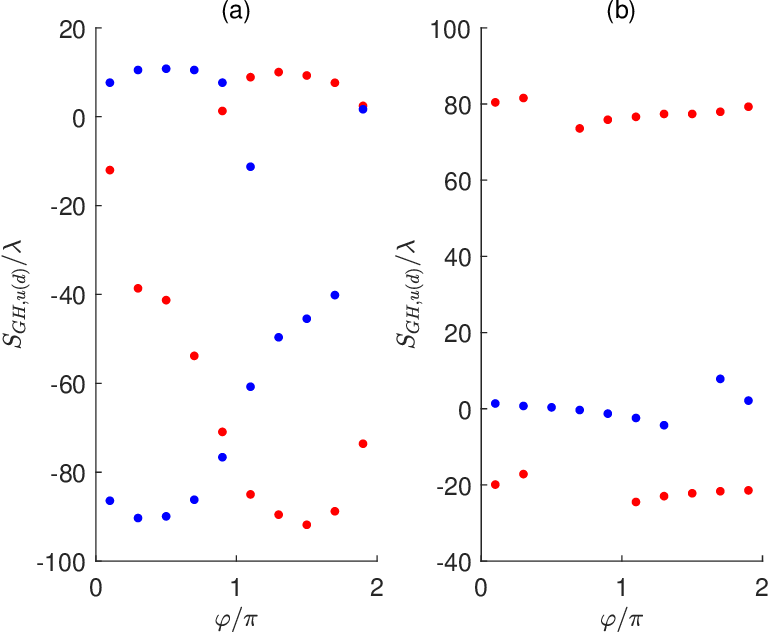}}
\caption{ Numerical value of GH shifts extracted from Fig. \ref{figure_slabGHt+all} and Fig. \ref{figure_slabGHt-all} are plotted in panel (a) and (b), respectively. The red and blue dots represent GH shifts of the outgoing beams to the up and down sides, respectively. }
\label{figure_slabGHphase}
\end{figure}

Numerical result of coherent GH shift in Fig. \ref{figure_slabGH} with complete set of $\varphi$ are plotted in Fig. \ref{figure_slabGHt+all} and Fig. \ref{figure_slabGHt-all}. The coherent GH shifts of the outgoing beams are extracted from the numerical results in Fig. \ref{figure_slabGHt+all} and Fig. \ref{figure_slabGHt-all} by calculating the distance between the black dots and the red (blue) dots at the same z, which are plotted in Fig. \ref{figure_slabGHphase}.

As $\theta_{Inc}=21.5^{o}$, $\varphi/\pi=0.3$, $0.5$, and $0.7$, the GH shifts of the outgoing beam to the up side are $-38.6\lambda$, $-41.3\lambda$, and $-53.8\lambda$, respectively, as shown in Fig. \ref{figure_slabGHphase}(a). Thus, GH shift versus $\varphi$ can be approximated as $S_{GH,u}\approx\lambda[-52.9+84.5\varphi/\pi-122.5(\varphi/\pi)^{2}]$. The sensitivity of the GH shift to $\varphi$ at $\varphi=\pi/2$ is given as $\frac{\partial S_{GH,u}}{\partial\varphi}|_{\varphi=\pi/2}=-0.2111\lambda$ $deg^{-1}$. The magnitude of the sensitivity is larger than that given by the stationary-phase method ($-0.1406\lambda$ $deg^{-1}$ as given in Sec. III A), because of the nonlinear relation between $S_{GH,u}$ and $\varphi$ for Gaussian beams' incidence.

As $\theta_{Inc}=21.5^{o}$, $\varphi/\pi=1.3$, $1.5$, and $1.7$, the GH shifts of the outgoing beam to the down side are $-49.6\lambda$, $-45.5\lambda$, and $-40.1\lambda$, respectively, as shown in Fig. \ref{figure_slabGHphase}(a). Thus, GH shift versus $\varphi$ can be approximated as $S_{GH,u}\approx\lambda[-44.6-25.0\varphi/\pi+16.3(\varphi/\pi)^{2}]$. The sensitivity of the GH shift to $\varphi$ at $\varphi=3\pi/2$ is $0.1319\lambda$ $deg^{-1}$, which is slight smaller than that given by the stationary-phase method ($+0.1406\lambda$ $deg^{-1}$ as given in Sec. III A).

As $\theta_{Inc}=-21.5^{o}$, $\varphi/\pi=1.3$, $1.5$, and $1.7$, the GH shifts of the outgoing beam to the up side are $77.0\lambda$, $77.4\lambda$, and $78.0\lambda$, respectively, as shown in Fig. \ref{figure_slabGHphase}(b). Thus, GH shift versus $\varphi$ can be approximated as $S_{GH,u}\approx\lambda[79.3-5.0\varphi/\pi+2.5(\varphi/\pi)^{2}]$. The sensitivity of the GH shift to $\varphi$ at $\varphi=3\pi/2$ is $0.0139\lambda$ $deg^{-1}$, which is much smaller than that given by the stationary-phase method ($+0.1426\lambda$ $deg^{-1}$ as given in Sec. III A). This is due to the large divergence angle of the incident Gaussian beams, which excites resonant modes outside of the resonant peak in Fig. \ref{figure_grateGH}(l).

\clearpage


\section*{References}

\begin{thebibliography}{99}


\bibitem{firstGH47} F. Goos and H. H$\ddot{a}$nchen, Ein neuer und fundamentaler Versuch zur Totalreflexion, Ann. Phys. 436, 333 (1947).

\bibitem{Brewster1} H. M. Lai and S. W. Chan, Large and negative Goos-H$\ddot{a}$nchen shift near the Brewster dip on reflection from weakly absorbing media, Opt. Lett. 27, 680 (2002).

\bibitem{Brewster2} L. Wang, H. Chen, and S. Zhu, Large negative Goos-H$\ddot{a}$nchen shift from a weakly absorbing dielectric slab, Opt. Lett. 30, 2936 (2005).

\bibitem{Brewster3} P. T. Leung, C. W. Chen, and H.-P. Chiang, Large negative Goos-H$\ddot{a}$nchen shift at metal surfaces, Opt. Commun. 276, 206(2007).


\bibitem{sppGH1} X. Yin, L. Hesselink, Z. Liu, N. Fang, and X. Zhang, Large positive and negative lateral optical beam displacements due to surface plasmon resonance, Appl. Phys. Lett. 85, 372(2004).

\bibitem{sppGH2} X. Yin and L. Hesselink, Goos-H$\ddot{a}$nchen shift surface plasmon resonance sensor, Appl. Phys. Lett. 89, 261108 (2006).

\bibitem{sppGH3} V. J. Yallapragada, A. P. Ravishankar, G. L. Mulay, G. S. Agarwal, and V. G. Achanta, Observation of giant Goos-H$\ddot{a}$nchen and angular shifts at designed metasurfaces, Sci. Rep. 6, 19319 (2016).

\bibitem{sppGH4} Y. Wang, Y. Hu, R. Xie, Q. Zeng, Y. Hong, X. Chen, P. Zhang, L. Zeng, Y. Zhang, S. Zeng, and H. Yang, Ultrasensitive label-free miRNA-21 detection based on MXene-enhanced plasmonic lateral displacement measurement, Nanophotonics 12, 4055 (2023).


\bibitem{fpcavityGH1} C.-F. Li and Q. Wang, Prediction of simultaneously large and opposite generalized Goos-H$\ddot{a}$nchen shifts for TE and TM light beams in an asymmetric double-prism configuration, Phys. Rev. E 69, 055601(R) (2004).

\bibitem{fpcavityGH2} L. Wang and S. Zhu, Giant lateral shift of a light beam at the defect mode in one-dimensional photonic crystals, Opt. Lett. 31, 101 (2006).

\bibitem{fpcavityGH3} K. V. Sreekanth, Q. Ouyang, S. Han, K.-T. Yong, and R. Singh, Giant enhancement of Goos-H$\ddot{a}$nchen shift at the singular phase of a nanophotonic cavity, Appl. Phys. Lett. 112, 161109 (2018).

\bibitem{fpcavityGH4} F. Wu, T. Liu, M. Luo, H. Li, and S. Xiao, Giant Goos-H$\ddot{a}$nchen shifts with high reflection driven by Fabry-Perot quasibound states in the continuum in double-layer gratings, Phys. Rev. B 109, 125411(2024).




\bibitem{blochGH1} I. V. Soboleva, V. V. Moskalenko, and A. A. Fedyanin, Giant Goos-H$\ddot{a}$nchen effect and Fano resonance at photonic crystal surfaces, Phys. Rev. Lett. 108, 123901 (2012).

\bibitem{blochGH2} Y. Wan, Z. Zheng, W. Kong, X. Zhao, Y. Liu, Y. Bian, and J. Liu, Nearly three orders of magnitude enhancement of Goos-H$\ddot{a}$nchen shift by exciting Bloch surface wave, Opt. Express 20, 8998 (2012).





\bibitem{tamnGH1} J. Tang, J. Xu, Z. Zheng, H. Dong, J. Dong, S. Qian, J. Guo, L. Jiang, and Y. Xiang, Graphene Tamm plasmon-induced giant Goos-H$\ddot{a}$nchen shift at terahertz frequencies, Chin, Opt. Lett. 17, 020007 (2019).

\bibitem{tamnGH2} J. Wu, F. Wu, K. Lv, Z. Guo, H. Jiang, Y. Sun, Y. Li, and H. Chen, Giant Goos-H$\ddot{a}$nchen shift with a high reflectance assisted by interface states in photonic heterostructures, Phys. Rev. A 101, 053838 (2020).

\bibitem{tamnGH3} Y. Ye, W. Chen, S. Wang, Y. Liu, and L. Jiang, Enhanced and tunable Goos-H$\ddot{a}$nchen effect of reflected light due to Tamm surface plasmons with Dirac semimetals, Results Phys. 43, 106105 (2022).




\bibitem{bicGH1} F. Wu, J. Wu, Z. Guo, H. Jiang, Y. Sun, Y. Li, J. Ren and H. Chen, Giant Enhancement of the Goos-H$\ddot{a}$nchen Shift Assisted by Quasibound States in the Continuum, Phys. Rev. Appl. 12, 014028(2019).

\bibitem{bicGH2} F. Wu, M. Luo, J. Wu, C. Fan, X. Qi, Y. Jian, D. Liu, S. Xiao, G. Chen, H. Jiang, Y. Sun and H. Chen, Dual quasibound states in the continuum in compound grating waveguide structures for large positive and negative Goos-H$\ddot{a}$nchen shifts with perfect reflection, Phys. Rev. A 104, 023518(2021).

\bibitem{bicGH3} M. Luo and F. Wu, Refractive-index sensing based on large negative Goos-H$\ddot{a}$nchen shifts of a wavy dielectric grating, Phys. Rev. Appl. 22, 014050(2024).




\bibitem{stationaryPt48} K. Artmann, Berechnung der Seitenversetzung des totalreflektierten Strahles, Ann. Phys. 437, 87 (1948).



\bibitem{raDongJianWen} Z.-P. Zhuang, H.-L. Zeng, X.-D. Chen, X.-T. He, and J.-W. Dong, Topological Nature of Radiation Asymmetry in Bilayer Metagratings, Phys. Rev. Lett. 132, 113801(2024).


\bibitem{bicItSelf1} C. W. Hsu, B. Zhen, A. D. Stone, J. D. Joannopoulos, and M. Solja$\check{c}$i$\acute{c}$, Bound state in the continuum, Nat. Rev. Mater. 1, 16048 (2016).

\bibitem{bicItSelf2} S. Joseph, S. Pandey, S. Sarkar, and J. Joseph, Bound states in the continuum in resonant nanostructures: An overview of engineered materials for tailored applications, Nanophotonics 10, 4175 (2021).

\bibitem{bicItSelf3} S. I. Azzam and A. V. Kildishev, Photonic bound states in the continuum: From basics to applications, Adv. Opt. Mater. 9, 2001469 (2021).

\bibitem{bicItSelf4} A. F. Sadreev, Interference traps waves in an open system: Bound states in the continuum, Rep. Prog. Phys. 84, 055901(2021).

\bibitem{bicVortex} B. Zhen, C.W. Hsu, L. Lu, A. D. Stone, and M. Solja$\check{c}$i$\acute{c}$, Topological nature of optical bound states in the continuum, Phys. Rev. Lett. 113, 257401(2014).




\bibitem{quasiBIC01} C. W. Hsu, B. Zhen, J. Lee, S. Chua, S. G. Johnson, J. D. Joannopoulos, and M. Solja$\check{c}$i$\acute{c}$, Observation of trapped light within the radiation continuum, Nature (London) 499, 188(2013).

\bibitem{quasiBIC02} Y. Yang, C. Peng, Y. Liang, Z. Li, and S. Noda, Analytical Perspective for Bound States in the Continuum in Photonic Crystal Slabs, Phys. Rev. Lett. 113, 037401 (2014).

\bibitem{quasiBIC03} K. Koshelev, S. Lepeshov, M. Liu, A. Bogdanov, and Y. Kivshar, Asymmetric Metasurfaces with High-Q Resonances Governed by Bound States in the Continuum, Phys. Rev. Lett. 121, 193903 (2018).

\bibitem{quasiBIC04} M. Minkov, I. A. D. Williamson, M. Xiao, and S. Fan, Zero-Index Bound States in the Continuum, Phys. Rev. Lett. 121, 263901 (2018).

\bibitem{quasiBIC05} J. Jin, X. Yin, L. Ni, M. Solja$\check{c}$i$\acute{c}$, B. Zhen, and C. Peng, Topologically enabled ultrahigh-Q guided resonances robust to out-of-plane scattering, Nature (London) 574, 501 (2019).

\bibitem{quasiBIC06} S. Li, C. Zhou, T. Liu, and S. Xiao, Symmetry-protected bound states in the continuum supported by all-dielectric metasurfaces, Phys. Rev. A 100, 063803 (2019).

\bibitem{quasiBIC07} S. Dai, P. Hu, and D. Han, Near-field analysis of bound states in the continuum in photonic crystal slabs, Opt. Express 28, 16288(2020).

\bibitem{quasiBIC08} B. Wang, W. Liu, M. Zhao, J. Wang, Y. Zhang, A. Chen, F. Guan, X. Liu, L. Shi, and J. Zi, Generating optical vortex beams by momentum-space polarization vortices centred at bound states in the continuum, Nat. Photonics 14, 623 (2021).

\bibitem{quasiBIC09} J. W. Yoon, S. H. Song, and R. Magnusson, Critical field enhancement of asymptotic optical bound states in the continuum, Sci. Rep. 5, 18301 (2015).

\bibitem{quasiBIC10} F. Monticone and A. Al$\grave{u}$, Bound states within the radiation continuum in diffraction gratings and the role of leaky modes, New J. Phys. 19, 093011 (2017).

\bibitem{quasiBIC11} Z. F. Sadrieva, I. S. Sinev, K. L. Koshelev, A. Samusev, I. V. Iorsh, O. Takayama, R. Malureanu, A. A. Bogdanov, and A. V. Lavrinenko, Transition from optical bound states in the continuum to leaky resonances: Role of substrate and roughness, ACS Photonics 4, 723 (2017).

\bibitem{quasiBIC12} H. M. Doeleman, F. Monticone, W. den Hollander, A. Al$\grave{u}$, and A. F. Koenderink, Experimental observation of a polarization vortex at an optical bound state in the continuum, Nat. Photonics 12, 397 (2018).

\bibitem{quasiBIC13} E. N. Bulgakov, D. N. Maksimov, P. N. Semina, and S. A. Skorobogatov, Propagating bound states in the continuum in dielectric gratings, J. Opt. Soc. Am. B 35, 1218 (2018).

\bibitem{quasiBIC14} S. I. Azzam, V. M. Shalaev, A. Boltasseva, and A. V. Kildishev, Formation of Bound States in the Continuum in Hybrid Plasmonic-Photonic Systems, Phys. Rev. Lett. 121, 253901(2018).

\bibitem{quasiBIC15} S. Joseph, S. Sarkar, S. Khan, and J. Joseph, Exploring the optical bound state in the continuum in a dielectric grating coupled plasmonic hybrid system, Adv. Opt. Mater. 9, 2001895(2021).

\bibitem{quasiBIC16} D. Liu, X. Yu, F. Wu, S. Xiao, F. Itoigawa, and S. Ono, Terahertz high-Q quasibound states in the continuum in laserfabricated metallic double-slit arrays, Opt. Express 29, 24779(2021).

\bibitem{quasiBIC17} Z. F. Sadrieva, M. A. Belyakov, M. A. Balezin, P. V. Kapitanova, E. A. Nenasheva, A. F. Sadreev, and A. A. Bogdanov, Experimental observation of a symmetric-protected bound state in the continuum in a chain of dielectric disks, Phys. Rev. A 99, 053804 (2019).

\bibitem{quasiBIC18} M. V. Rybin, K. L. Koshelev, Z. F. Sadrieva, K. B. Samusev, A. A. Bogdanov, M. F. Limonov, and Y. S. Kivshar, High-Q Supercavity Modes in Subwavelength Dielectric Resonators, Phys. Rev. Lett. 119, 243901 (2017).

\bibitem{quasiBIC19} E. Melik-Gaykazyan, K. Koshelev, J. Choi, S. S. Kruk, A. Bogdanov, H. Park, and Y. Kivshar, From Fano to quasi-BIC resonances in individual dielectric nanoantennas, Nano Lett. 21, 1765(2021).

\bibitem{quasiBIC20} H. Hemmati and R. Magnusson, Resonant dual-grating metamembranes supporting spectrally narrow bound states in the continuum, Adv. Opt. Mater. 7, 1900754 (2019).

\bibitem{quasiBIC21} L. Cong and R. Singh, Symmetry-protected dual bound states in the continuum in metamaterials, Adv. Opt. Mater. 7, 1900383(2019).




\bibitem{cPointFromV1} X. Yin, J. Jin, M. Solja$\check{c}$i$\acute{c}$, C. Peng, and B. Zhen, Observation of topologically enabled unidirectional guided resonances, Nature (London) 580, 467 (2020).

\bibitem{cPointFromV2} Y. Zeng, G. Hu, K. Liu, Z. Tang, and C.-W. Qiu, Dynamics of topological polarization singularity in momentum space, Phys. Rev. Lett. 127, 176101 (2021).

\bibitem{cPointFromV3} X. Yin, T. Inoue, C. Peng, and S. Noda, Topological unidirectional guided resonances emerged from interband coupling, Phys. Rev. Lett. 130, 056401 (2023).


\bibitem{bicSHG1} X. Tu, S. Feng, J. Li, Y. Xing, F. Wu, Tingtin. Liu, and S. Xiao, Enhanced second-harmonic generation in high-Q all-dielectric metasurfaces with backward frequency conversion, Phys. Rev. A 109, 063522 (2024).

\bibitem{bicSHG2} G. Zhang, K. Zhang, X. Huang, X. Chen, X. Wang, W. Tao, and Y. Mao, Highly robust and efficient second harmonic generation device based on lithium niobate, Phys. Rev. B 111, 155302 (2025).


\bibitem{theFEM1} M. Luo and Q. H. Liu, Extraordinary transmission of a thick film with a periodic structure consisting of strongly dispersive materials, J. Opt. Soc. Am. B 28, 629 (2011).

\bibitem{theFEM2} M. Luo and Q. H. Liu, Three-dimensional dispersive metallic photonic crystals with a bandgap and a high cutoff frequency, J. Opt. Soc. Am. A 27, 1878 (2010).

\bibitem{theFEM3} M. Luo, Q. H. Liu, and J. Guo, A spectral element method calculation of extraordinary light transmission through periodic subwavelength slits, J. Opt. Soc. Am. B 27, 560(2010).

\bibitem{theFEM4} M. Luo and Q. H. Liu, Accurate determination of band structures of 2D dispersive anisotropic photonic crystals by the spectral element method, J. Opt. Soc. Am. A 26, 1598(2009).

\bibitem{theFEM5} M. Luo and Q. H. Liu, Spectral element method for band structures of three-dimensional anisotropic photonic crystals, Phys. Rev. E 80, 056702 (2009).

\bibitem{theFEM6} M. Luo, Q. H. Liu, and Z. Li, Spectral element method for band structures of two-dimensional anisotropic photonic crystals, Phys. Rev. E 79, 026705 (2009).



\bibitem{tcmt01} S. Fan, W. Suh, and J. D. Joannopoulos, J. Opt. Soc. Am. A 20, 569 (2003).

\bibitem{tcmt02} F. Wu, C. Fan, K. Zhu, J. Wu, X. Qi, Y. Sun, S. Xiao, H. Jiang, and H. Chen, Phys. Rev. B 105, 245417(2022).

\bibitem{tcmt03} M. Luo and F. Wu,  Phys. Rev. A 106, 063514(2022).




\end{thebibliography}
\end{document}